\documentclass[a4paper,11pt]{article}
\usepackage{jinstpub} % for details on the use of the package, please see the JINST-author-manual
\usepackage{lineno}
\usepackage[amssymb]{SIunits}
\usepackage[T1]{fontenc}
\usepackage{subcaption}
\usepackage{xcolor}
\usepackage{comment}

\title{\boldmath Hybrid microchannel plate PMT characterisation on a test-beam using a Ring Imaging Cherenkov setup}

\author[c]{J.~Alozy}
\author[c]{R.~Ballabriga}
\author[a]{N.~V.~Biesuz}
\author[a,b]{R.~Bolzonella }
\author[c]{M.~Campbell}
\author[a]{G.~Cavallero}
\author[a,b]{V.~Cavallini}
\author[a]{A.~Cotta~Ramusino}
\author[a,b]{M.~Fiorini}
\author[a,b]{E.~Franzoso}
\author[a,b]{M.~Guarise}
\author[c]{X.~Llopart Cudie}
\author[d,e,1]{G.~Romolini \note{Corresponding author.}}
\author[a,b]{A.~Saputi}
\author[b]{D.~Vincenzi}

\affiliation[a]{INFN Ferrara, Via Saragat 1, 44122 Ferrara, Italy}
\affiliation[b]{Università degli Studi di Ferrara, Via Saragat 1, 44122 Ferrara, Italy}
\affiliation[c]{CERN, 1211 Geneva 23, Switzerland}
\affiliation[d]{INFN Bari,  Via E. Orabona 4, 70125, Bari, Italy}
\affiliation[e]{Università degli Studi di Bari, Via E. Orabona 4, 70125, Bari, Italy}

\emailAdd{gabriele.romolini@cern.ch}

\abstract{
A novel photodetector based on a vacuum tube with microchannel plate and encapsulated CMOS ASIC has been tested at the CERN SPS high energy hadron beam, using a Ring Imaging Cherenkov setup. The tube was operated at a very low gain of 9.9$\times$10$^3$ and allowed the imaging of single-photons from Cherenkov rings with timing resolution of about 280~ps and with 55~$\mathrm{\mu m}$ anode pixel pitch.
}


\begin{document}
\maketitle
\flushbottom

\section{Introduction}\label{sec:introduction}
A novel detector for of single optical photons counting has been developed within the 4DPHOTON project \cite{Fiorini_2018, Alozy_2022} based on a vacuum tube design incorporating a transmission photocathode, a microchannel plate (MCP) for electron multiplication, and the Timepix4 ASIC \cite{Llopart_2022} used as a pixelated anode.
This hybrid MCP-PMT architecture combines the excellent timing performance and high gain provided by the MCP with the fine spatial granularity and fully digital, event-driven readout of the Timepix4 ASIC. Compared to conventional MCP-PMTs, the pixelated anode significantly enhances spatial resolution and reduces event ambiguities under high-rate conditions, while preserving the fast timing characteristics of the MCP stage.
\begin{figure}[htbp]
\centering
    \centering
    \includegraphics[width=.8\textwidth]{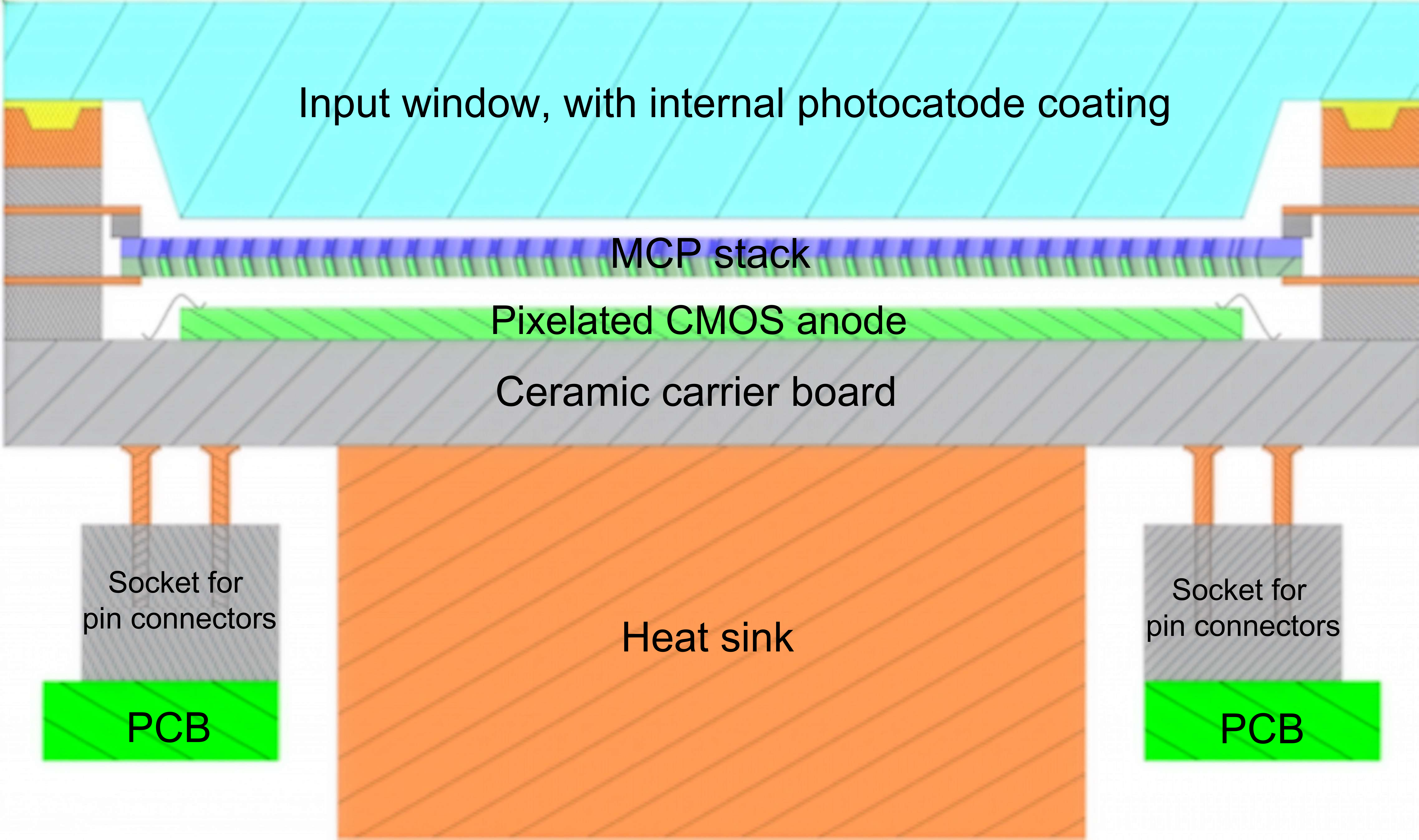}
    \caption{Simplified cross section of the detector assembly.}
    \label{fig:4DPHOTON}
\end{figure}

A test-beam was performed at the CERN SPS North Area to characterise the performance of the first 4DPHOTON detector prototypes, the detector under test (DUT), operating in a Ring-Imaging Cherenkov (RICH) configuration, shown in Figure~\ref{fig:4DPHOTON}. 
The main goal of the test-beam was to demonstrate the feasibility of a RICH detector based on the hybrid MCP-PMT technology. The data collected in the test-beam also allow to give an estimate of the timing resolution achievable under realistic conditions.

The tested prototype was produced by Hamamatsu Photonics with a stack of 3 MCPs and 1d end-spoiling, and was one of the first prototypes available at the time of the test-beam. The photodetector has an active area of approximately 28$\times$25 mm$^2$ and anode pixel pitch of 55 $\mu$m.

\section{Experimental setup} \label{Sec:test_beam_setup}

The test-beam has been performed at the H8 beamline of the CERN SPS, which provided a secondary hadron beam with 150 GeV$/c$ momentum, with about $80\%$ of protons, $20\%$ of pions, $1\%$ of muons and a negligible fraction of electrons.  A typical beam intensity of $10-100\kilo$ particles for each spill of $4.8\,\second$ was used.
 
\begin{figure}[htbp]
\centering
\subfloat[]
{
    \centering
    \hspace{-0.75cm}\includegraphics[width=.8\textwidth]{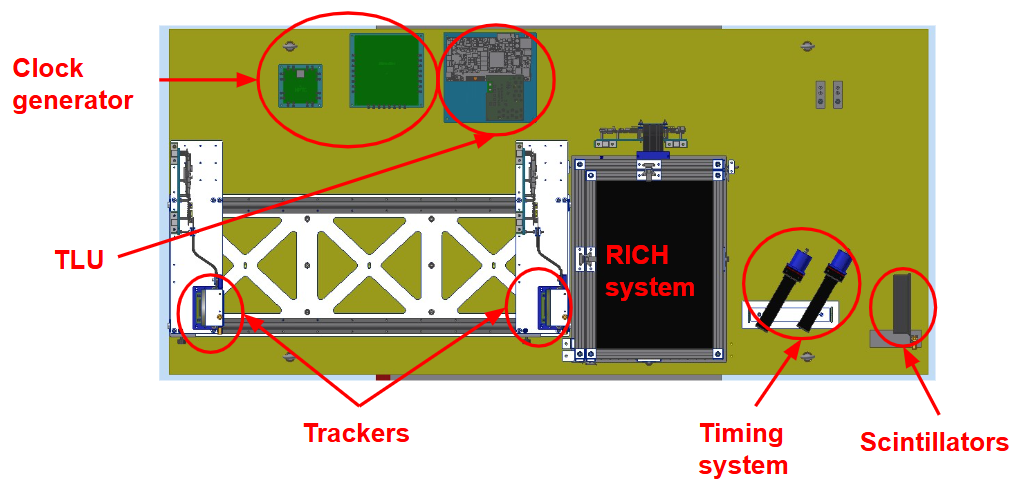}
    \label{fig:testbeam_setup_above}
}
\vspace{0.5cm}
\subfloat[]
{
    \centering
    \includegraphics[width=.6\textwidth]{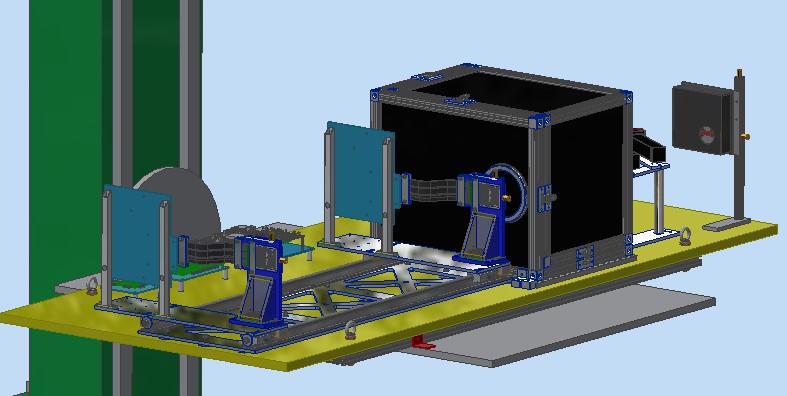}
    \label{fig:test_beam_setup_side}
}
\caption{\textbf{(a)} Schematic drawing of the test-beam setup (top view): it comprises a tracking station, a RICH system, a time reference system and two scintillators for beam alignment. An external clock generator and a Trigger Logic Unit (TLU) are used to provide an external reference clock and external T0\_sync and shutter signals to the three Timepix4. \\
\textbf{(b)} 3D CAD rendering of the experimental apparatus, mounted on a movable table.}
\label{fig:test_beam_setup}
\end{figure}

The experimental setup is presented in Fig.~\ref{fig:test_beam_setup}. It consists of three main components: a tracking system, a Ring-Imaging Cherenkov (RICH) system and timing reference detectors for precise event time measurement. The tracking system consists of two assemblies, each comprising a Timepix4 ASIC bump-bonded to a 300~$\mu$m-thick pixelated silicon sensor. Referred to as tracker 1 and tracker 2, they are positioned upstream of the experimental setup, with tracker 1 and tracker 2 located farther from and closer to the DUT, respectively. The detectors are spaced $100\,\centi\meter$ apart to achieve high precision in reconstructing particles trajectories, thanks to the 55~$\mu $m sensor pixel pitch.
The RICH system, enclosed within a light-tight box downstream of the tracking system, comprises a plano-convex LA4545 lens (f=100~mm), with its back metallized that serves both as a Cherenkov radiator and as the initial focusing element. A LA4078 lens (f=75.3~mm) is paired with the radiator lens and is used to further focus the photons. Photons generated inside this second lens are not focused to the DUT. Simulation studies have shown that less than one photon per event is expected. These two lenses are separated by an air gap of $100~\mu$m to enable photon reflection from the first lens. A tilted planar mirror made of aluminium, with the side blackened, reflects photons perpendicularly to the beam axis.
After the mirror, three LA4795 lenses (f=200~mm) are placed, followed by a LA4384 lens (f=90.3~mm). The DUT follows this optical system, and is placed at a distance of 180 mm from the center of the mirror, in order to create a circular image of the Cherenkov ring with a diameter of approximately 17~mm.
All the lenses are made of fused silica with a refractive index in the range of [1.45, 1.54] in the useful Cherenkov wavelength range, from 300 to 800 nm.
The simulated optical system is shown in Fig.~\ref{fig:geant4_optics_system}.
The timing reference system is placed downstream of the RICH system along the beam direction, and is composed of a pair of plexiglass parallelepiped-shaped bars that act as Cherenkov radiators, tilted with an angle of 65$^{\degree}$ with respect to the beam axis, and optically coupled to standard single-anode MCP-PMT tubes, that provide fast timing measurements as a reference for evaluating the DUT timing resolution.
A pair of scintillators have also been used to monitor the beam intensity and align the setup to the beam.

The detectors in the system were synchronised by using  a shared common clock running at $40~$MHz.
Data acquisition was initiated by using external timing signals from the SPS fed to a custom Trigger Logic Unit (TLU).
This board, running  synchronously with the $40~$MHz shared clock,  samples the warning-of-extraction (WE) signal from SPS and generates a shutter signal for each spill.
The delay between the WE signal from the SPS and the shutter signal and the duration of the shutter signal are configurable ensuring that the data acquisition occurred only during spills.
The shutter signal was then split using a high precision 1:8 LVDS buffer and forwarded to all the detectors in the system enabling data-acquisition.
On reception of this signal the data acquisition systems forwards data from each detector to local servers using standard UDP-IP frames on 10G ethernet connections, depending on the detector type.
At the local server level, data was handled by multiple threads, allowing for simultaneous storage and redirection to the control room computer for monitoring.
Data was automatically backed-up to a local storage server and mirrored to the CERN EOS system, providing redundancy without interrupting acquisition. Overall, the software infrastructure enabled synchronised and stable operation of the multiple Timepix4 detectors, while offering immediate feedback to optimise the data-taking strategy.

The C++ based DataPix4 framework~\cite{Cavallini:2025uwz} was used to manage both trackers and the DUT, control data acquisition and monitor the data stream in real time.
 DataPix4  also served as online monitor by performing online clustering  by aggregating hits belonging to the same photon event in real-time, which was instrumental to assess data quality during  data taking.

  The silicon sensors of the trackers were biased at 100~V. For the DUT, a voltage difference of 2100~V was applied to the MCP stack, while  voltage differences of $V_{\text{PC-MCP}}=75~\text{V}$ and $V_{\text{MCP-TPX4}}=150~\text{V}$ were applied between the photocathode and top MCP, and between the bottom MCP and the Timepix4.  Since large clusters can cause localised peaks in the ASIC power consumption, potentially leading to excessive voltage drops, affecting the signals transmission to or from the ASIC, an MCP bias voltage of 2100 V was selected for the test-beam measurements to ensure stable data taking, although the MCP stack had been successfully tested in the laboratory at voltages of up to 2400 V. The DUT was constantly cooled using copper heat sink in thermal contact to the input window and the ceramic carrier board back-side, as shown in Figure~\ref{fig:4DPHOTON}, setting a temperature of $-10^{\degree}\text{C}$ in the thermostatic bath of a recirculating chiller.

\begin{figure}[htbp]
\centering
\subfloat[]
{
    \centering
    \includegraphics[width=.475\textwidth]{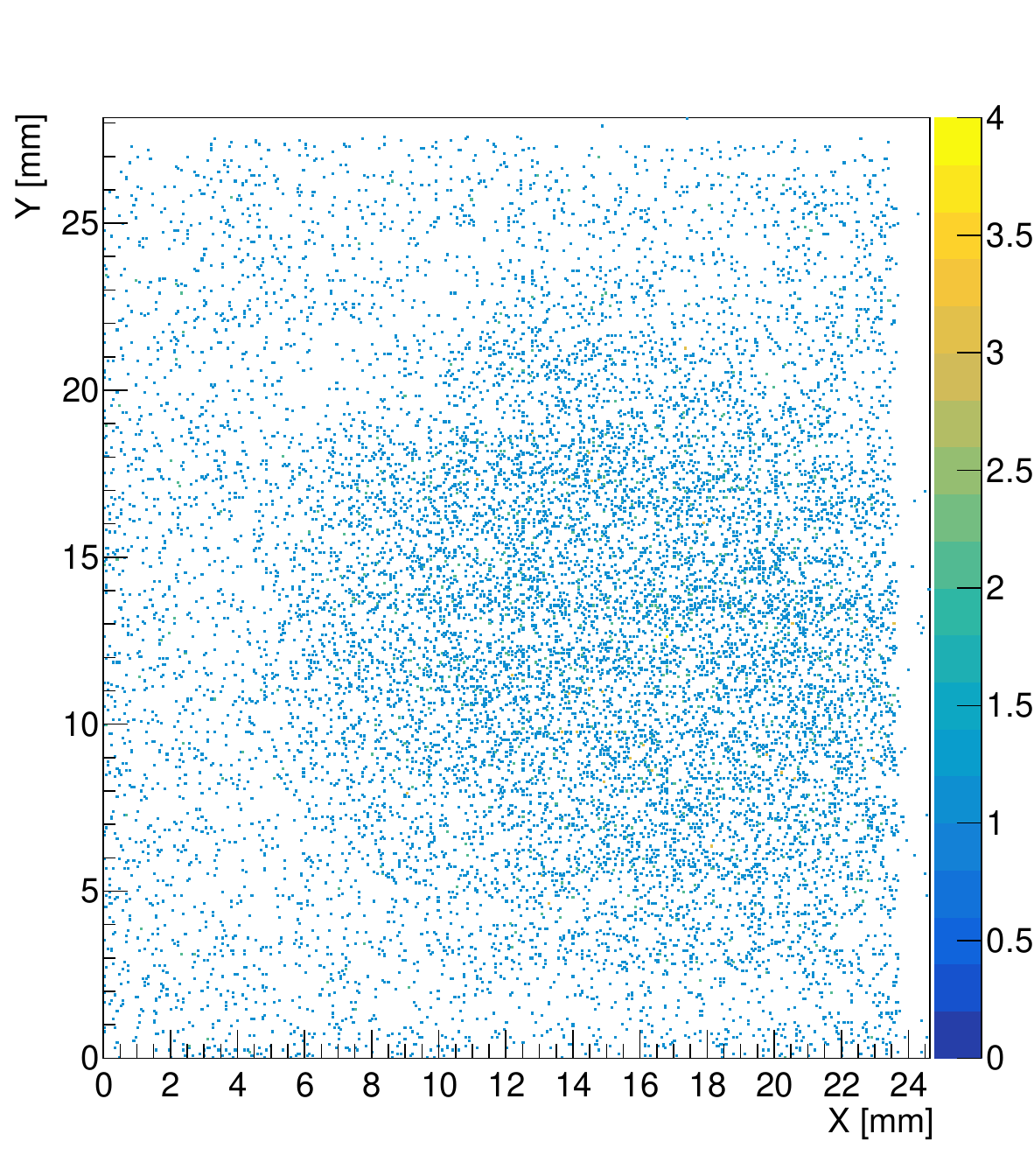}
    \label{fig:XY1}
}
\subfloat[]
{
    \centering
    \includegraphics[width=.475\textwidth]{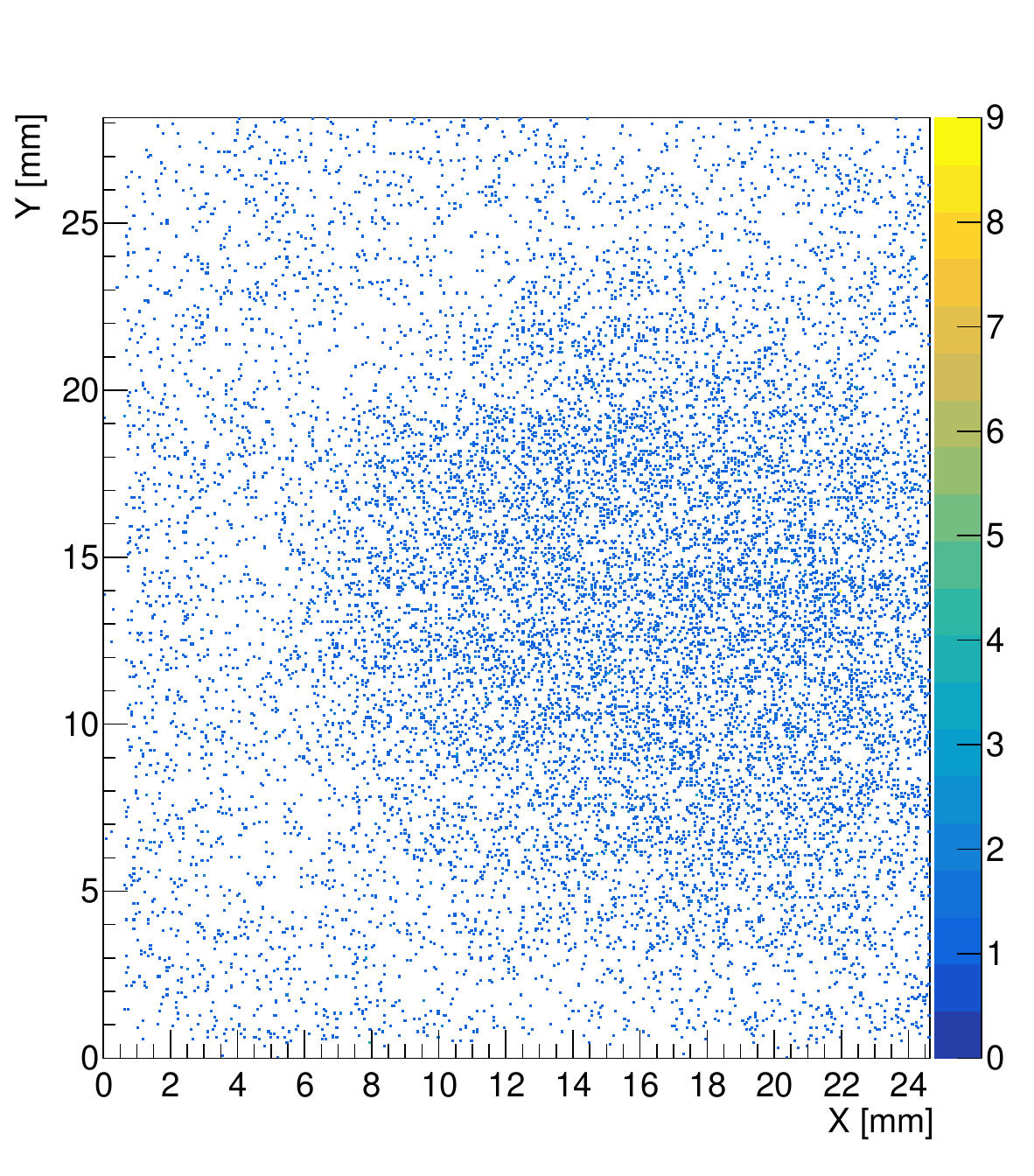}
    \label{fig:XY2}
}
\caption{Two-dimensional spatial distribution of the beam on the first \textbf{(a)}  and second  \textbf{(b)} tracking station.}
    \label{fig:BeamProfile}
\end{figure}

\begin{figure}
    \centering
    \includegraphics[width=0.85\linewidth]{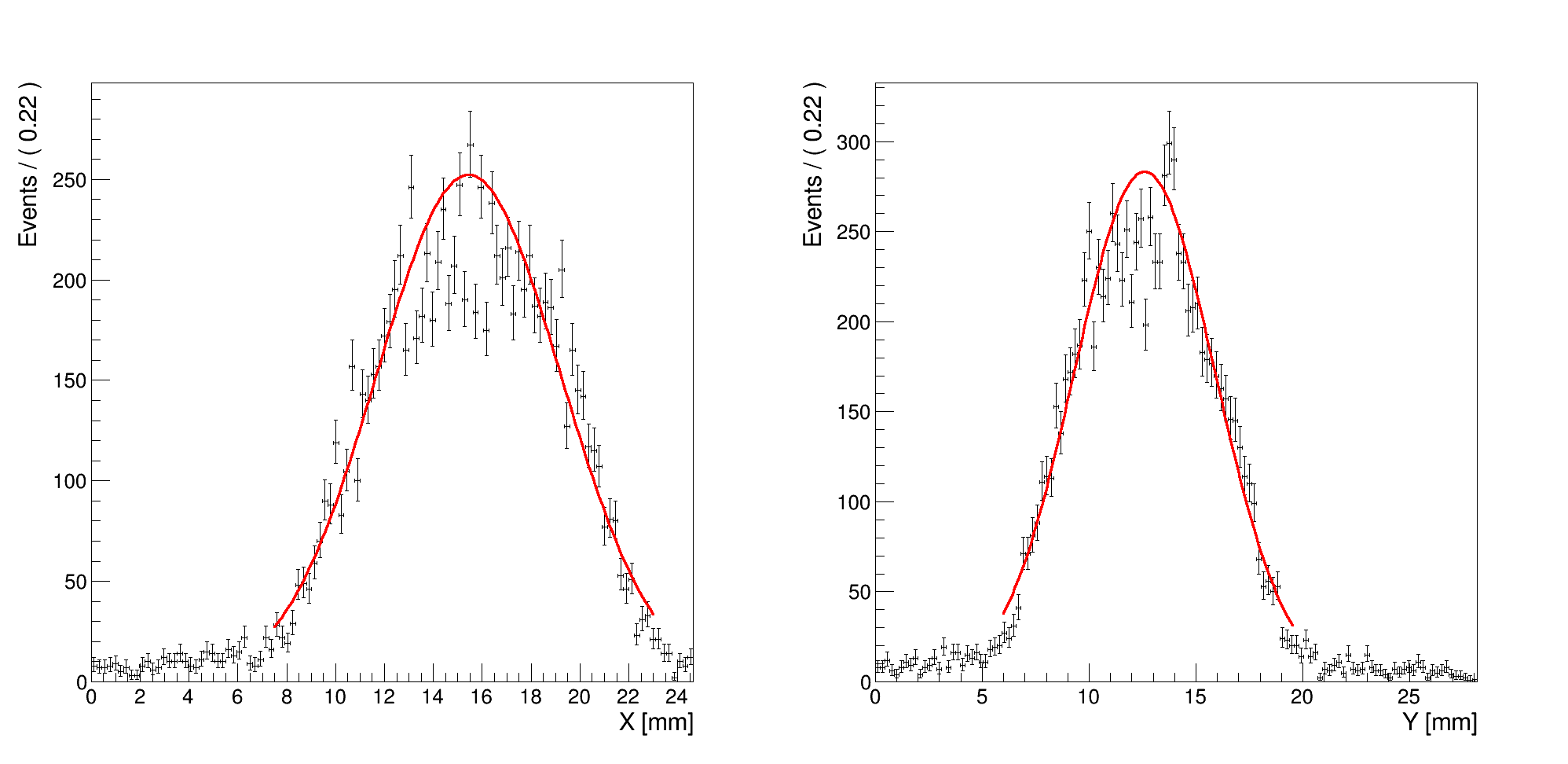}
    \includegraphics[width=0.85\linewidth]{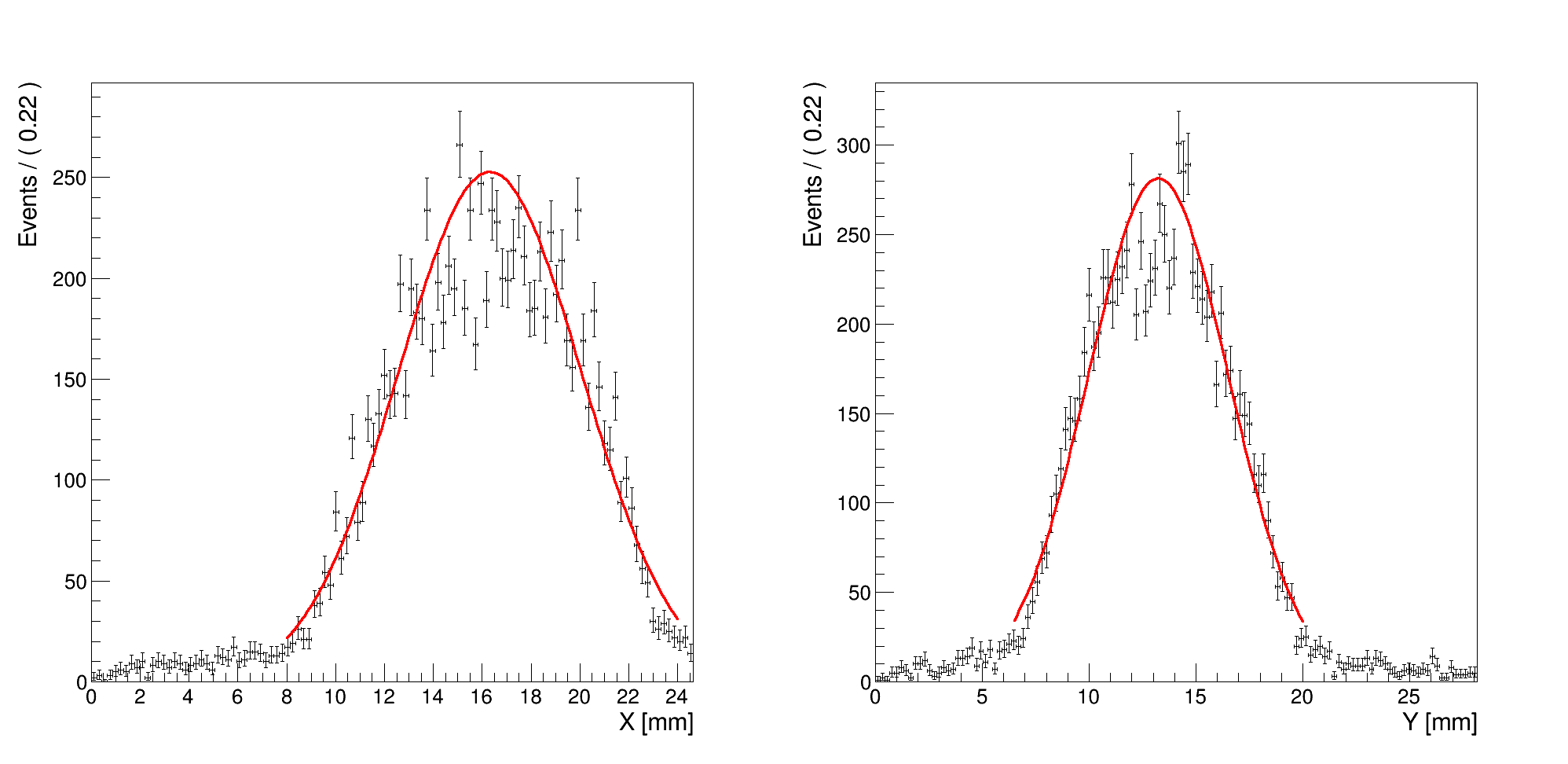}
    \caption{Tracker~1 (top) and  tracker~2 (bottom) $x$ (left) and  $y$ (right) projected hit distribution from data collected over 50 spills.  Gaussian fits are superimposed.}
    \label{fig:TrackerProjections}
\end{figure}

 The beam spot was measured to be gaussian on both perpendicular directions to the beam-line, with $\sigma_x = 3.76\pm0.04$~mm and $\sigma_y= 3.28\pm0.04$~mm and divergence compatible with $\sim200$~$\mu$rad, as measured by the tracking detectors. Fig.~\ref{fig:BeamProfile} shows the beam spatial distribution perpendicular to the beam axis, while Fig.~\ref{fig:TrackerProjections} shows the corresponding projected distrubutions.
 
\section{Simulation} \label{Subsec:Testbeam_simulation}

A complete simulation was developed using the Geant4 \cite{Geant4_1} toolkit to model the entire experimental setup and to provide accurate predictions of the photon timing and spatial distributions. 
The simulation includes the beam and its interactions with the trackers, the Cherenkov radiator and the  photon production, the transport of photons through the optical system, as well as the beam interaction with the Cherenkov detectors used for timing reference.
The simulation takes into account several key factors, including photocathode quantum efficiency, angular and energy distribution of the particles and the open-area-ratio of the MCP, fixed to $40\%$ in the simulation. After the test-beam, the simulation has been updated using the beam parameters extracted from the trackers.

This simulation tool was employed to estimate the expected Cherenkov ring resolution, analyse the timing distribution of photons within a Cherenkov ring, and was used in the design phase to find the optimal detector configuration by comparing different setup options.

\begin{figure}[h!]
\centering
    \includegraphics[width=.75\textwidth]{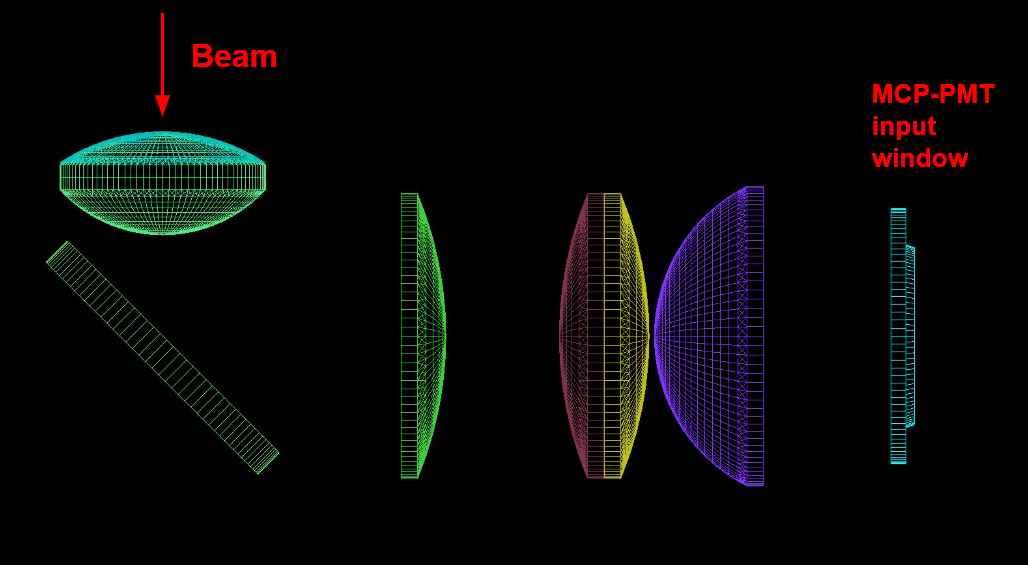}
    \caption{Geant4 simulation of the complete optical system of the RICH setup.}
    \label{fig:geant4_optics_system}
\end{figure}

\begin{figure}[h!]
\centering
    \includegraphics[width=.8\textwidth]{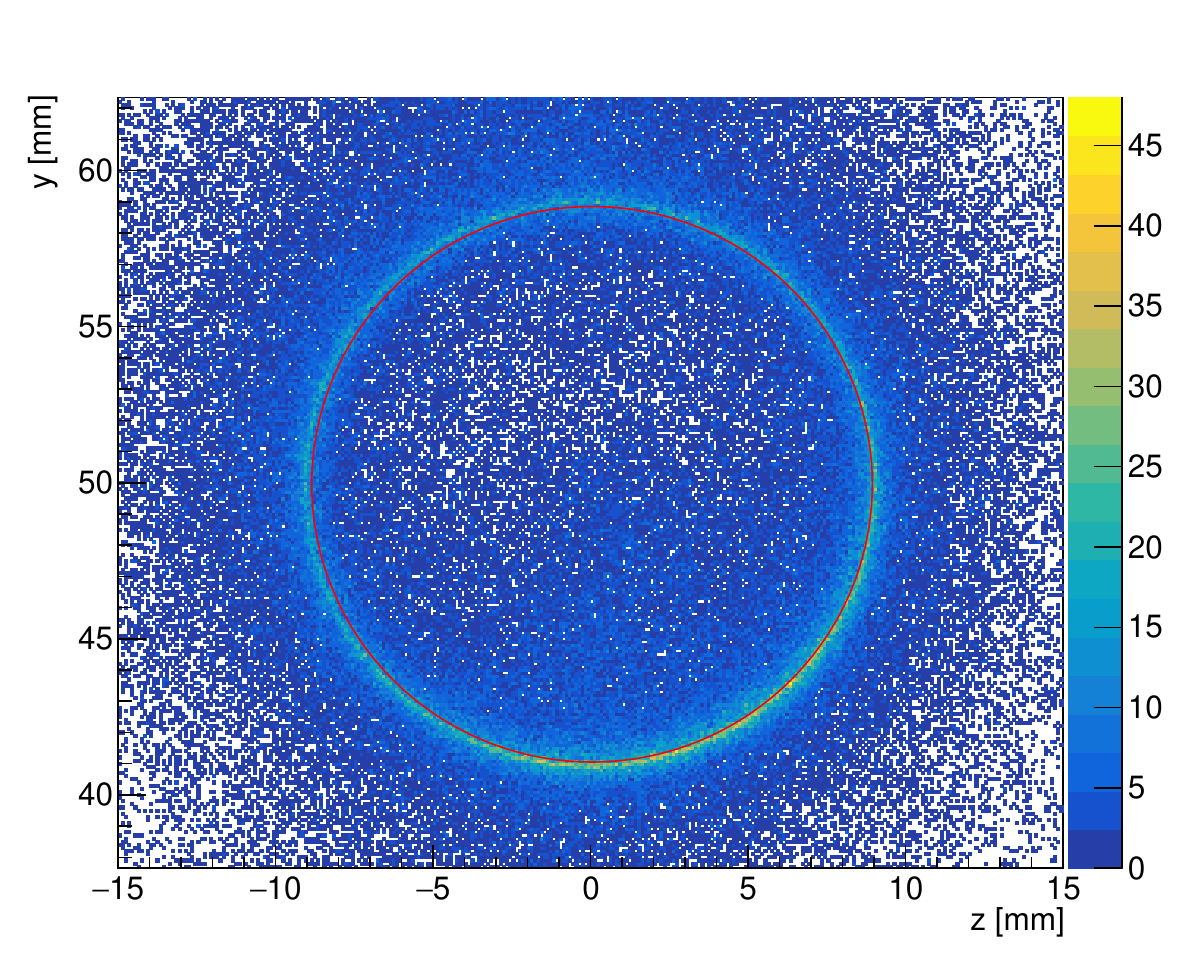}
    \caption{Simulated Cherenkov photons distribution on the detector prototype. Superimposed a circular fit (red line).}
    \label{fig:SimRing}
\end{figure}

The optical elements simulated by Geant4 are shown in Fig.~\ref{fig:geant4_optics_system}. 
Photons produced in a typical event are focused in a ring-shaped pattern. The photons distribution obtained from a thousand simulated events is reported in Fig.~\ref{fig:SimRing}, the superimposed fit measured a radius of $8.69\pm0.01$~mm. The small asymmetry obeserved in the $y$ axis is due to the small beam divergence.

\begin{figure}[h!]
\centering
    \includegraphics[width=0.75\textwidth]{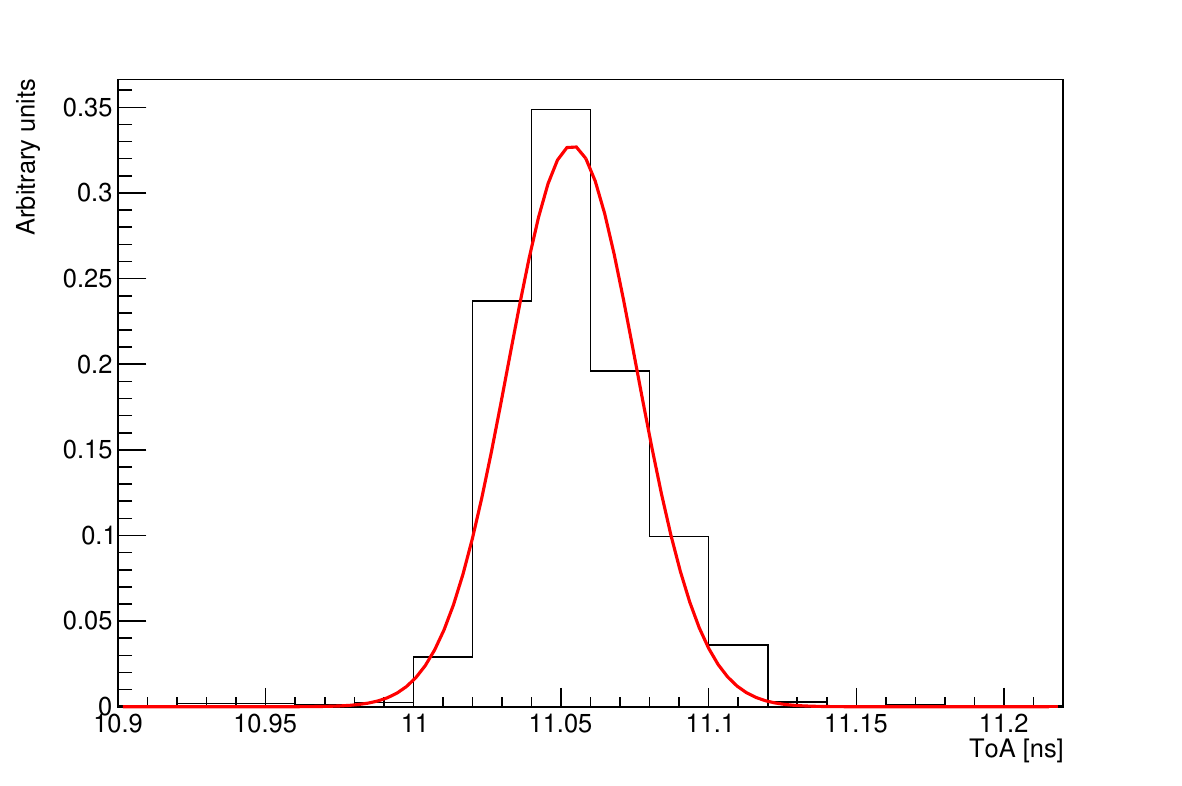}
    \caption{Simulated time of arrival distribution of the Cherenkov photons on the detector with superimposed Gaussian fit~($\sigma$ = 24~ps).}
    \label{ToASim}
\end{figure}

The arrival-time distribution of photons belonging to a single Cherenkov event has a standard deviation of approximately $20~$ps, as can be seen from Fig.~\ref{ToASim}. The simulation does not account for effects related to the detector, such as dark counts or finite time-bin size.


\section{Methodology and test-beam results} \label{Sec:Experimental_results}

\begin{figure}[h!]
\centering
    \includegraphics[width=.85\textwidth]{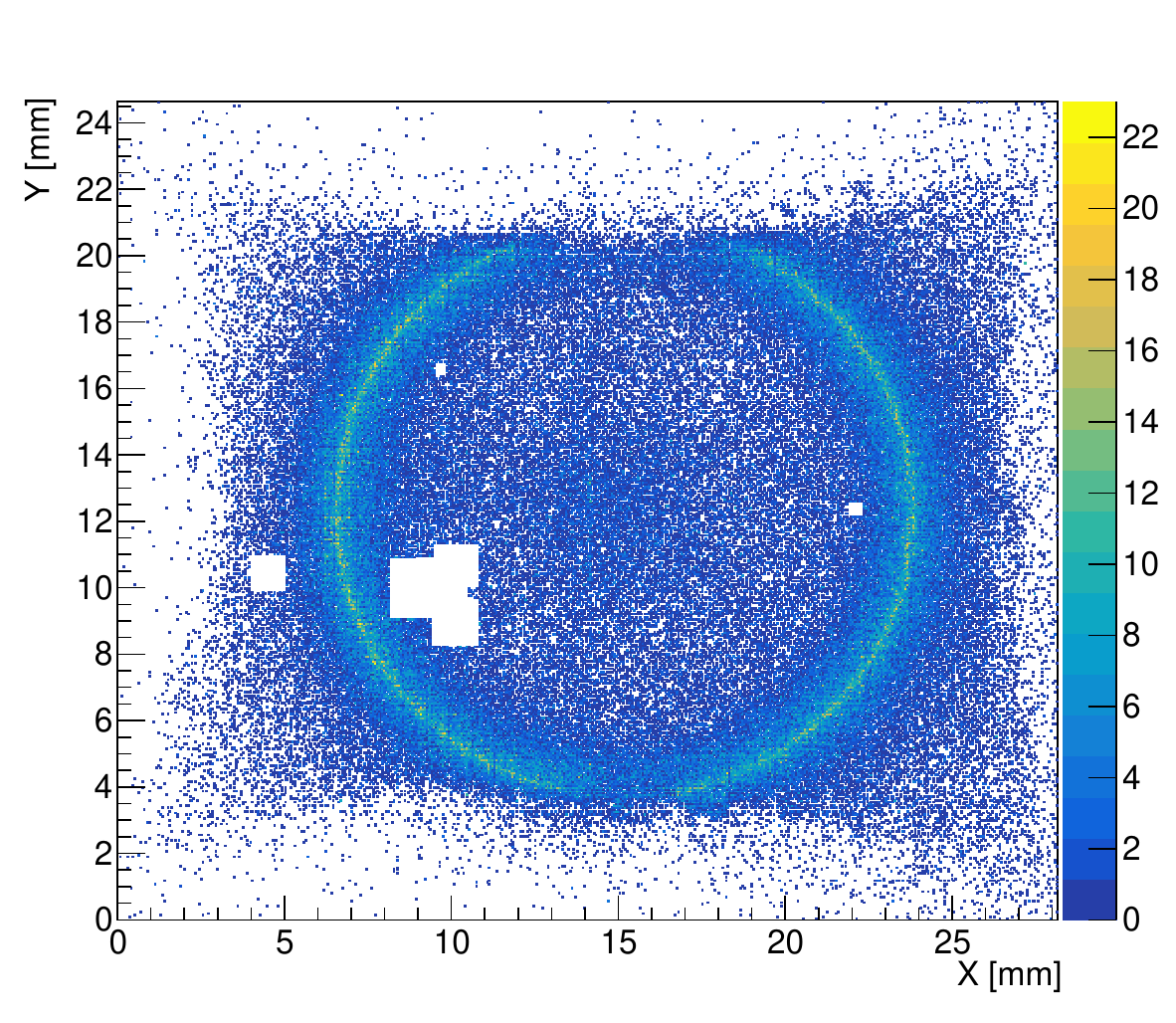}
    \caption{Cumulative distribution on the detector prototype of 50 spills. The blank rectangular regions correspond to noisy areas of the MCP that were masked.}
    \label{fig:Cumulativering}
\end{figure}

No hardware triggers were employed during data acquisition, except for the shutter signal at the start of the each spill.
A candidate Cherenkov ring is accepted only if the two tracking detectors register hits within a 25~ns time window. In addition, every photon in the DUT must occur within 50~ns of the hit recorded by the tracker that is closest to the RICH system. Furthermore, the $x$ and $y$ coordinates of the tracks measured by the trackers are required to be within $1\sigma$ of the corresponding measured beam spot coordinates. This cut is applied in order to select proton passing near the center of the lenses and therefore produce Cherenkov photons that are correctly focused on the DUT.
Figure~\ref{fig:Cumulativering} presents the cumulative distribution of the clusters on the DUT from data collected over 50 spills. 
Charge-calibration and timing corrections, including time-walk effects, are applied using the same method described in~\cite{Bolzonella:2024gdf}. The Cherenkov ring appears with the expected size and position. A non-uniform background pattern is also visible. This background is consistent with photons not correctly focused by the lens system, as can be seen in Fig.~\ref{fig:SimRing} and also dark counts. The shape is due to gain non-uniformitites, consistent with what observed during laboratory characterisation.

\begin{figure}[h!]
\centering
    \includegraphics[width=.75\textwidth]{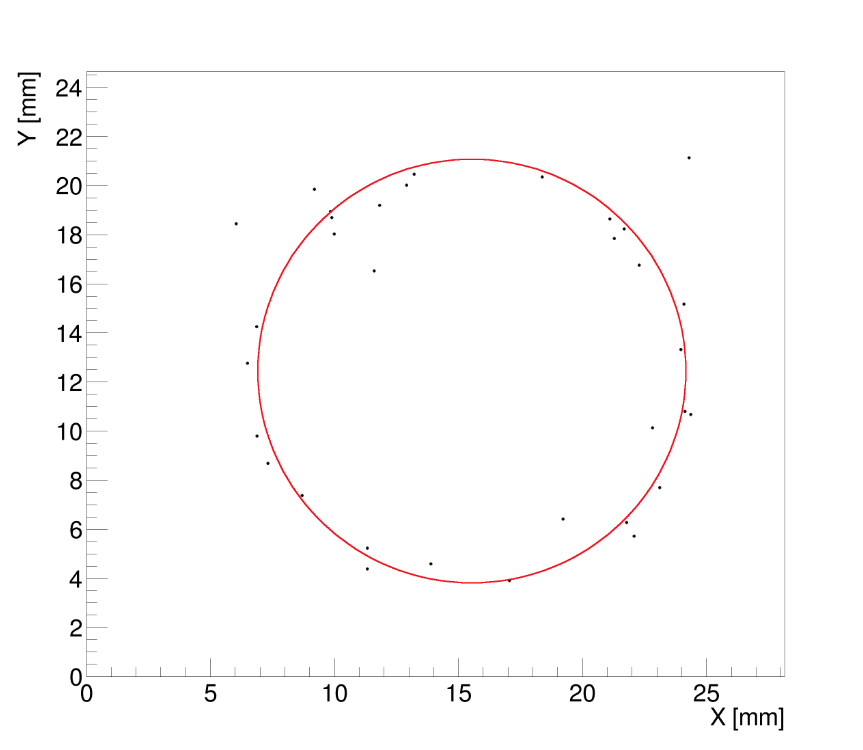}
    \caption{Example of photon hits distribution for a single ring candidate, with superimposed fitted ring.}
    \label{fig:Singlefit}
\end{figure}

\begin{figure}[htbp]
\centering
\subfloat[]
{
    \centering
    \includegraphics[width=.475\textwidth]{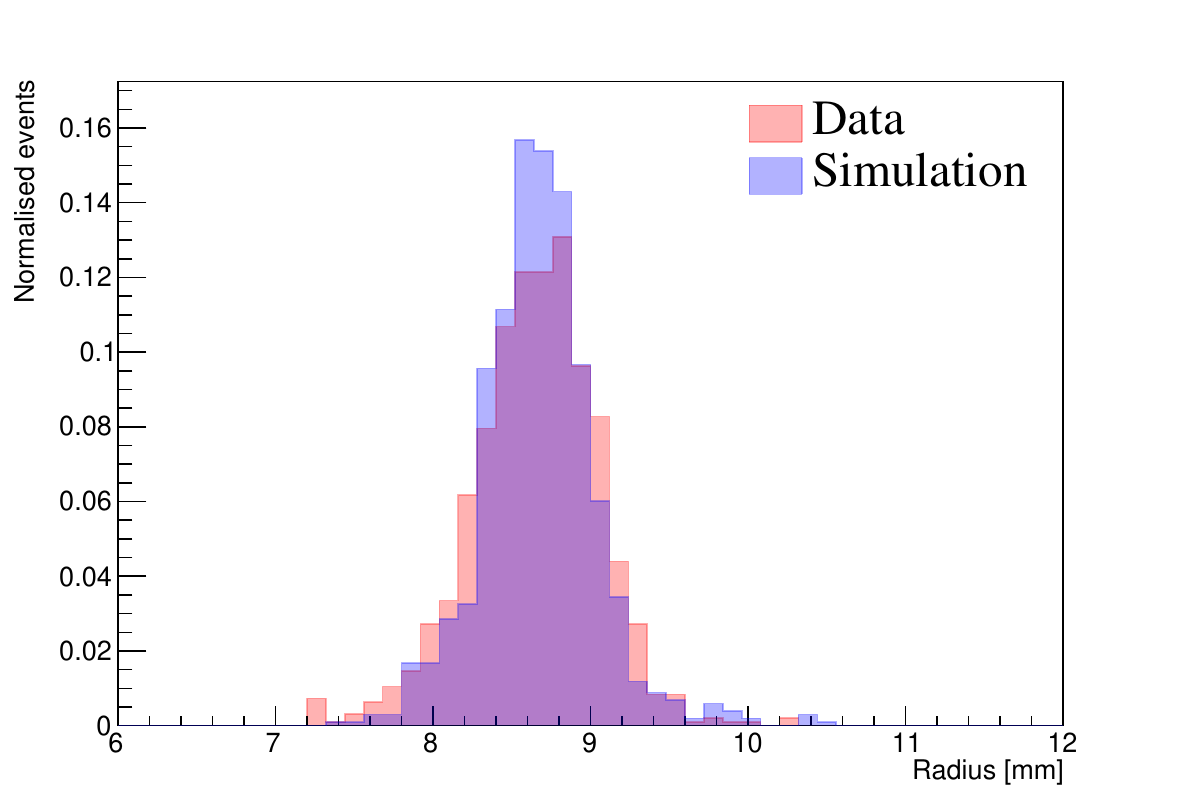}
    \label{fig:ComparisonRing}
}
\subfloat[]
{
    \centering
    \includegraphics[width=.475\textwidth]{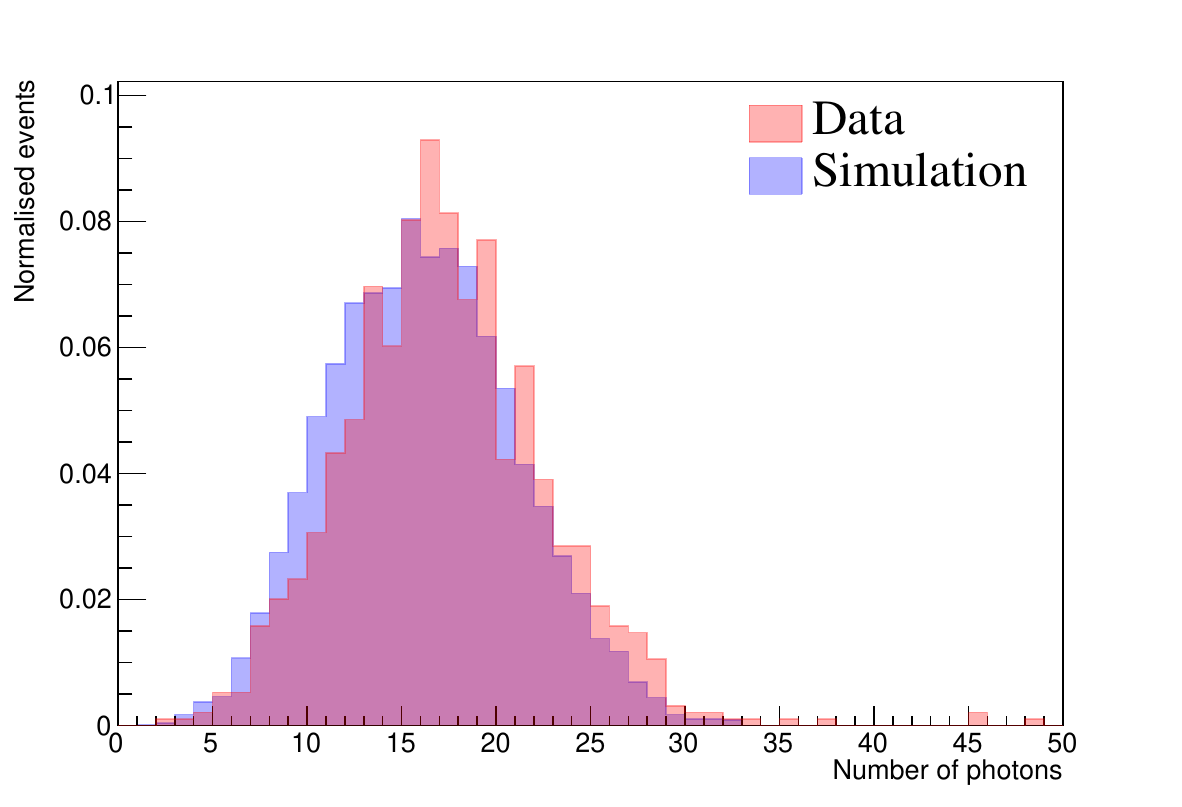}
    \label{fig:ComparisonNPhot}
}

    \caption{\textbf{(a)} Measured distributions of reconstructed Cherenkov rings radii, compared to simulation. \textbf{(b) }Distribution of the number of photons per Cherenkov ring reconstructed from data, compared with simulation.}
    \label{fig:Comparison}
\end{figure}

A ring fit procedure has been developed in order to extract the Cherenkov ring radius and compare it with the simulation result.
Each ring candidate is fitted independently, and the ring centre coordinates ($x, y$) and radius are free parameters of the fit. For each photon in the candidate the distance between photon and ring is minimised. Only fit candidates for which the minimisation procedure converged successfully are considered. The percentage of events with a non-convergent fit is 5$\%$. A fit for a single event is shown in Figure~\ref{fig:Singlefit}. Distribution of the reconstructed radii extracted from data, and compared with simulation, is shown in Figure~\ref{fig:ComparisonRing}. 
The mean of the distribution measured from data is $R=8.64\pm0.01$~mm, while the value extracted from simulation is $R_{\mathrm{sim}}=8.67\pm0.01$~mm. The two values are compatible with each other within $3\sigma$, considering statistical uncertainties only.

The  photon counts distribution has been further analysed to estimate the number of detected Cherenkov photons. This estimation is based on the results of the previously described single-event fit. Only photons located within 1~mm of the fitted Cherenkov ring are considered in the analysis. Figure~\ref{fig:ComparisonNPhot} presents the resulting distribution of the number of photons per event, together with the corresponding distribution from the Geant4 simulation.
The mean number of photons per ring extracted from data is measured to be $17\pm1$, compatible with simulation $16 \pm 1$ .

\begin{figure}[h!]
\centering
    \includegraphics[width=.85\textwidth]{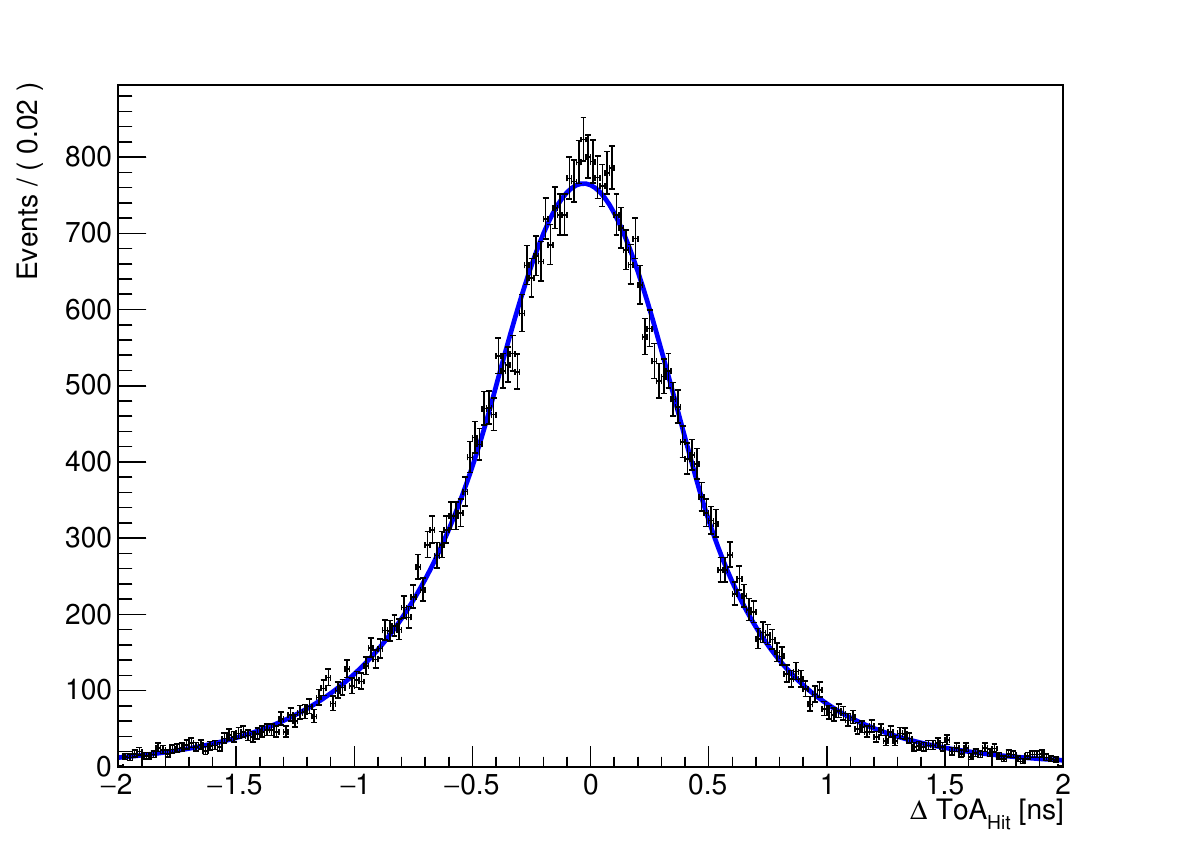}
    \caption{Distribution of the time of arrival difference of the photons with the fit result distribution.}
    \label{fig:TimingFit}
\end{figure}

The initial strategy to measure the  DUT time resolution was to use the pair of Cherenkov fast timing detectors as reference. During data taking, issues were identified in the timing reference system, that prevented the use of those data in the analysis. Due to the low bias voltage applied, and the pixel sensor planar geometry with no gain, the trackers cannot be used either as a time reference, since their resolution was measured to be in the order of 1~$\nano\second$.
Therefore, to estimate the DUT time resolution, only the timing information from the candidate Cherenkov rings themselves are used.
For each ring candidate, the associated photons are randomly divided into two subsets. Each photon in the first subset is then paired with a photon randomly extracted from the second subset. The difference of time of arrival for each pair is computed. This procedure suppresses effects that are common to all photons in a given Chereckov ring candidate, such as global geometric biases.
An unbinned maximum likelihood fit has been applied to the distribution obtained from the data, using a Crystal Ball function~\cite{Skwarnicki:1986xj} to take into account possible tails in the distribution due to asymmetries. Figure~\ref{fig:TimingFit} shows the time difference distribution with the fit result superimposed. 
The timing resolution obtained from the fit is $\sigma_{\text{timing}} = 401 \pm 7$~ps\, corresponding to the width of the gaussian core of the distribution. The full width at half maximum is evaluated to be 978$\pm 6~$ps from the resulting fit function.
Since this value is derived from the distribution of the difference between two statistically independent subsets, the intrinsic resolution is given by dividing the measured value by $\sqrt{2}$. The measured single photon time resolution is then $\sigma_{\text{timing}} = 283 \pm 5$~ps.

The time resolution was also studied as a function of the number of photons
associated to the ring candidate.
The extracted resolution was found not independent from the photon number, at low number of photons the measured resolution is increased by $\sim25\%$.
Since the single-photon resolution is expected to be independent of
the number of photons, this behaviour suggests that the sample of hits associated to a
ring candidate contains a component that is not made of Cherenkov photons
produced by the beam particle.
The origin of this contamination has not been established.
The value $\sigma_{\text{timing}} = 401 \pm 7$~ps, obtained from the full sample, is dominated by candidates with a large number of photons and is compatible with the plateau value.

The timing resolution measured during the test-beam is mainly limited by electronics: the low gain of the MCP stack at the bias voltage of 2100~V and the charge spread over a number of pixels,  reduces the charge of the signal at the preamplifier input and gives large jitter and time walk~\cite{Heijhoff:2022ifa}.

\begin{figure}[!h]
\centering
\subfloat[]
{
    \centering
    \includegraphics[width=.475\textwidth]{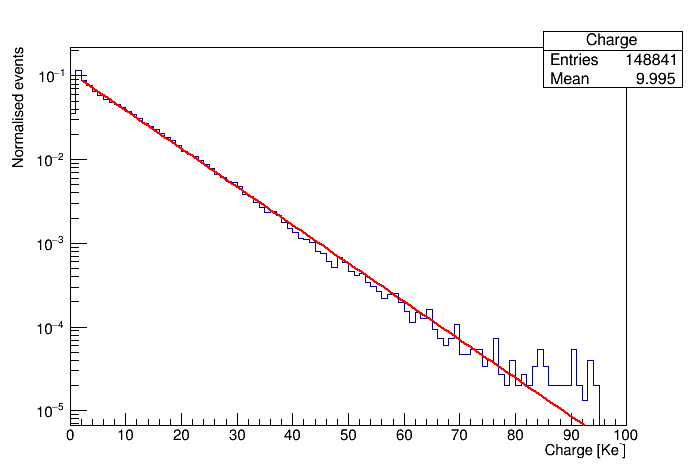}
    \label{Clusterinfos_charge}
}
\subfloat[]
{
    \centering
    \includegraphics[width=.475\textwidth]{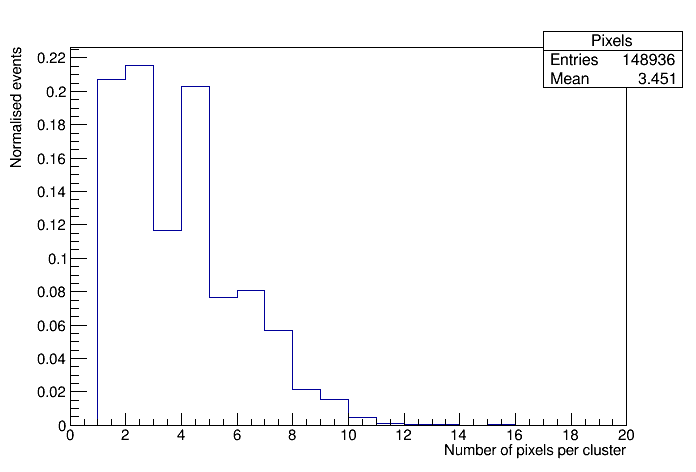}
    \label{Clusterinfos_pixel}
}

\subfloat[]
{
    \centering
    \includegraphics[width=.475\textwidth]{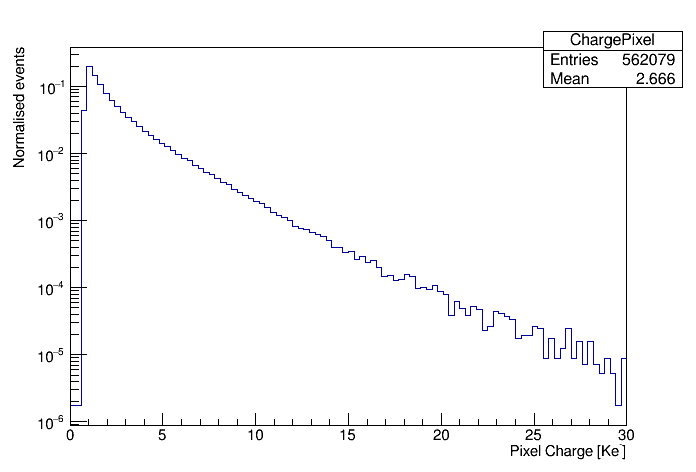}
    \label{Clusterinfos_pixelCharge}
}
    \caption{\textbf{(a)} Cluster charge distribution of reconstructed photons. The mean of the distribution, corresponding to the gain, is ($9.9 \pm 0.1$) ke-. \textbf{(b)} Cluster size distribution of reconstructed photons. \textbf{(c)} Pixel charge distribution of reconstructed photons. The mean of the distribution is ($2.7\pm0.2$) ke-.}
    \label{Clusterinfos}
\end{figure}

Figure~\ref{Clusterinfos_charge} shows the cluster charge distribution, whose mean value gives us a gain of $9.9 \pm 0.1$ k$e^-$, and it is superimposed to a decreasing exponential distribution, whose fit gives a comparable result, 9.9$\pm$0.2 k$e^-$.
Figure~\ref{Clusterinfos_pixel} represents the cluster size distribution, that shows that typically the charge cloud multiplied by the MCP stack is shared by 3.5 pixels on average.
Finally, Figure~\ref{Clusterinfos_pixelCharge} shows the charge distribution per pixel, which is sensed by the preamplifier circuit. The average amounts to 2.7 k$e^-$, which is a very low value: the Timepix4 analog front-end has a typical jitter of about 300~ps for that charge~\cite{Heijhoff:2022ifa}, which must be added to the TDC resolution and to the sensor contribution.

\section{Conclusions}\label{sec:Conclusions}

The results of the test-beam performed at the CERN SPS using the first prototype of the novel hybrid photodetector have been presented. 
Using a solid Cherenkov radiator and a focusing system, we were able to image complete Cherenkov rings onto the active area of a single detector.
The experimental results show good agreement with the simulations, including the expected number of detected Cherenkov photons and Cherenkov rings radius.
Detector operation at the very low gain of $9.9 \pm 0.1$ k$e^-$ was demonstrated. The detector operated reliably for the whole duration of the test-beam (1 week) and a time resolution of $\sigma_{\text{timing}} = (283 \pm 5)$~ps was measured, limited by the analog front-end jitter for low gain signals.

\newpage

\appendix

\acknowledgments

This work was carried out in the context of the Medipix4 Collaboration based at CERN, and in the framework of the MEDIPIX4 project funded by INFN CSN5.
This project has received funding from the European Research Council (ERC) under the European Union's Horizon 2020 research and innovation programme (Grant agreement No. 819627, 4DPHOTON project).
We express our gratitude to the SPS team for providing the beam at the H8 line and for their support during the test-beam campaign, as well as to Thomas Schneider for the lens metallisation.

\bibliographystyle{JHEP}
\bibliography{biblio.bib}

\end{document}